# Neutron Spectra from Intermediate-Energy Nucleus-Nucleus Reactions

Hiroshi Iwase[*], Yoshiyuki Iwata[†], Takashi Nakamura[**], Konstantin Gudima[‡], Stepan Mashnik[§], Arnold Sierk[§] and Richard Prael[§]

[*]*GSI, Planckstr. 1, 64291 Darmstadt, Germany*
[†]*National Institute of Radiological Sciences, 4-9-1 Anagawa, Inage, Chiba 263-8555, Japan*
[**]*Tohoku University, Aoba, Aramaki, Aoba-ku, Sendai 980-8579, Japan*
[‡]*Institute of Applied Physics, Academy of Science of Moldova, Chişinău, MD-2028, Moldova*
[§]*Los Alamos National Laboratory, Los Alamos, NM 87545, USA*

**Abstract.** Double-differential cross sections of neutron production at angles from 0 to 110 degrees from many reactions induced by light and medium nuclei on targets from $^{12}$C to $^{208}$Pb, at several incident energies from 95 to 600 MeV/nucleon have been measured recently at the Institute of Physical and Chemical Research (RIKEN) Ring Cyclotron in Japan and at the Heavy-Ion Medical Accelerator of the National Institute of Radiological Science in Chiba, Japan using the time-of-flight technique. We have analyzed all these new measurements using the Quantum Molecular Dynamics (QMD) model, the Oak Ridge intranuclear cascade model HIC, the ISABEL intranuclear cascade model from LAHET, and the Los Alamos version of the Quark-Gluon String Model code LAQGSM03. On the whole, all four models used here describe reasonably well most of the measured neutron spectra, although different models agree differently with data from specific reactions and some serious discrepancies are observed for some reactions. We present here some illustrative results from our study, discuss possible reasons for some of the observed discrepancies and try to outline ways to further improve the tested codes in order to address these problems.

## INTRODUCTION

Recently, intermediate-energy heavy ions have been used in various fields of nuclear physics and medical application, especially cancer therapy. Since 1994, more than 2000 patients were treated with $^{12}$C beam at the Heavy-Ion Medical Accelerator HIMAC at the National Institute of Radiological Science (NIRS) in Chiba, Japan. At the therapy pilot project at GSI Darmstadt (Gesellschaft fuer Schwerionenforschung) Germany, more than 250 patients were treated with $^{12}$C beam since 1997 and a new clinical-based facility is under construction in Heidelberg, Germany. Several institutes in the world have started or planned to build radioactive beam facilities where intermediate-energy radioactive heavy ions are to be used for investigating exotic nuclei far off the line of stability and so on. To design these facilities, radiation shielding is essential to protect workers and nearby inhabitants from penetrating neutrons produced by heavy ions. Data on double-differential neutron cross sections from such reactions are indispensable in estimating radiation sources for accelerator shielding design.

Reactions with intermediate-energy heavy ions are also of interest to astrophysics and data for such reactions are very important to estimate the radiation shielding needed to protect astronauts, computers, and other electronics on spacecraft.

Recently, double-differential cross sections of neutron production at angles from 0 to 110 degrees from many reactions induced by low- and medium-mass nuclei on targets from $^{12}$C to $^{208}$Pb, at several incident energies from 95 to 600 MeV/nucleon have been measured at RIKEN and HIMAC using the time-of-flight technique [1, 2]. We have analyzed these measurements using the Japan version of the Quantum Molecular Dynamics (QMD) model coupled with the statistical decay model in the code JQMD [3], the Oak Ridge intranuclear cascade model HIC by Bertini et al. [4] followed by a standard evaporation calculation with EVAP-4 [5], the ISABEL intranuclear cascade model [6] followed by the Dresner evaporation model code [5] from LAHET [7], and the Los Alamos version of the Quark-Gluon String Model [8], contained in the code LAQGSM03 [9].

## RESULTS AND DISCUSSION

The measurements are described in [1, 2] and details of the models used here can be found in [3-9]. Some of our results are presented in Figs. 1–3. More results and a detailed description of our work will be published later.

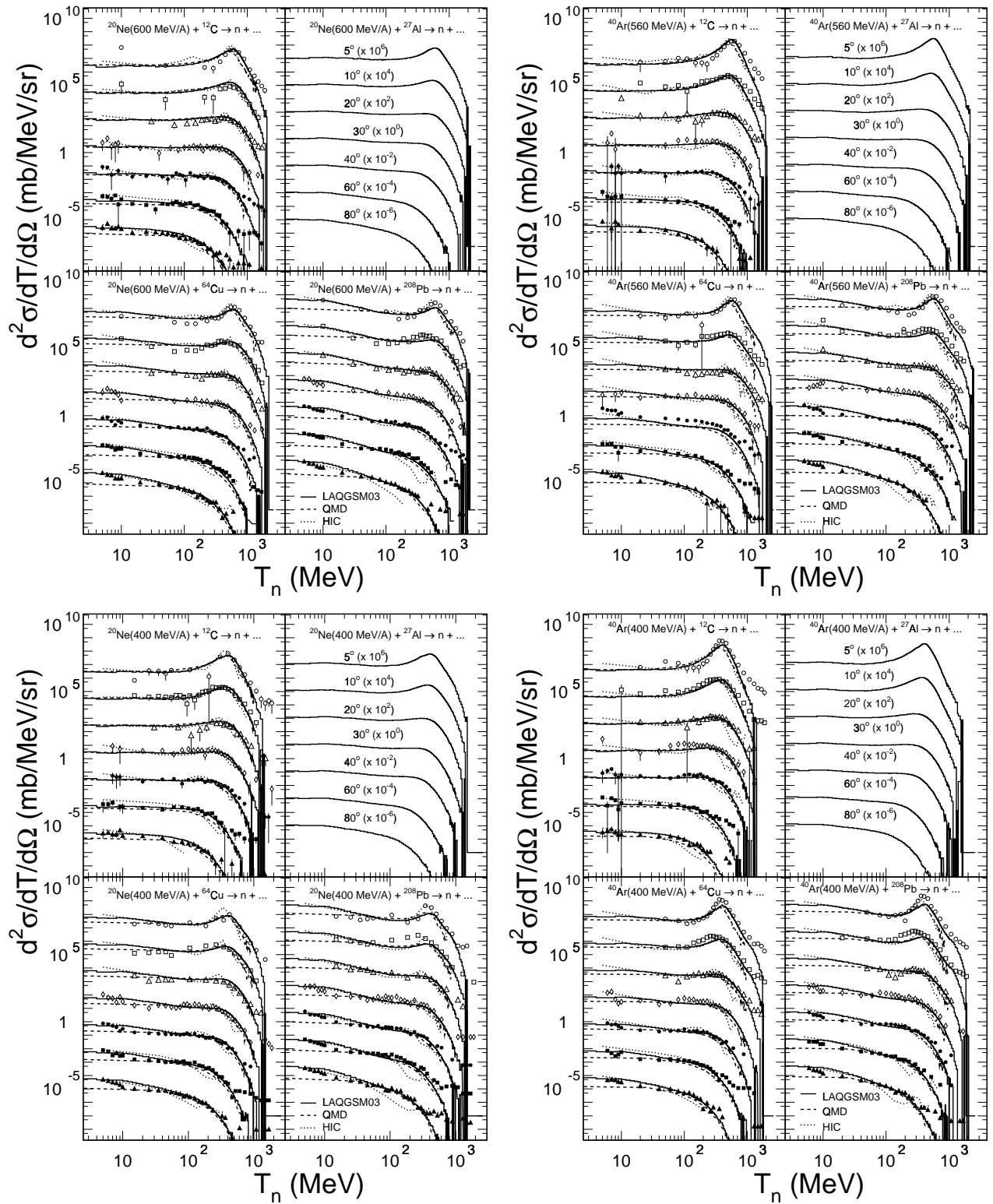

**FIGURE 1.** Comparison of measured [2] double differential cross sections of neutrons from 600 MeV/nucleon $^{20}$Ne, 560 MeV/nucleon $^{40}$Ar, 400 MeV/nucleon $^{20}$Ne, and 400 MeV/nucleon $^{40}$Ar on $^{12}$C, $^{27}$Al, $^{64}$Cu, and $^{208}$Pb with our present LAQGSM03 [9] results and calculations by JQMD [3] and HIC [4]. Experimental data for these reactions on Al are not yet available so we present here only our predictions from LAQGSM03.

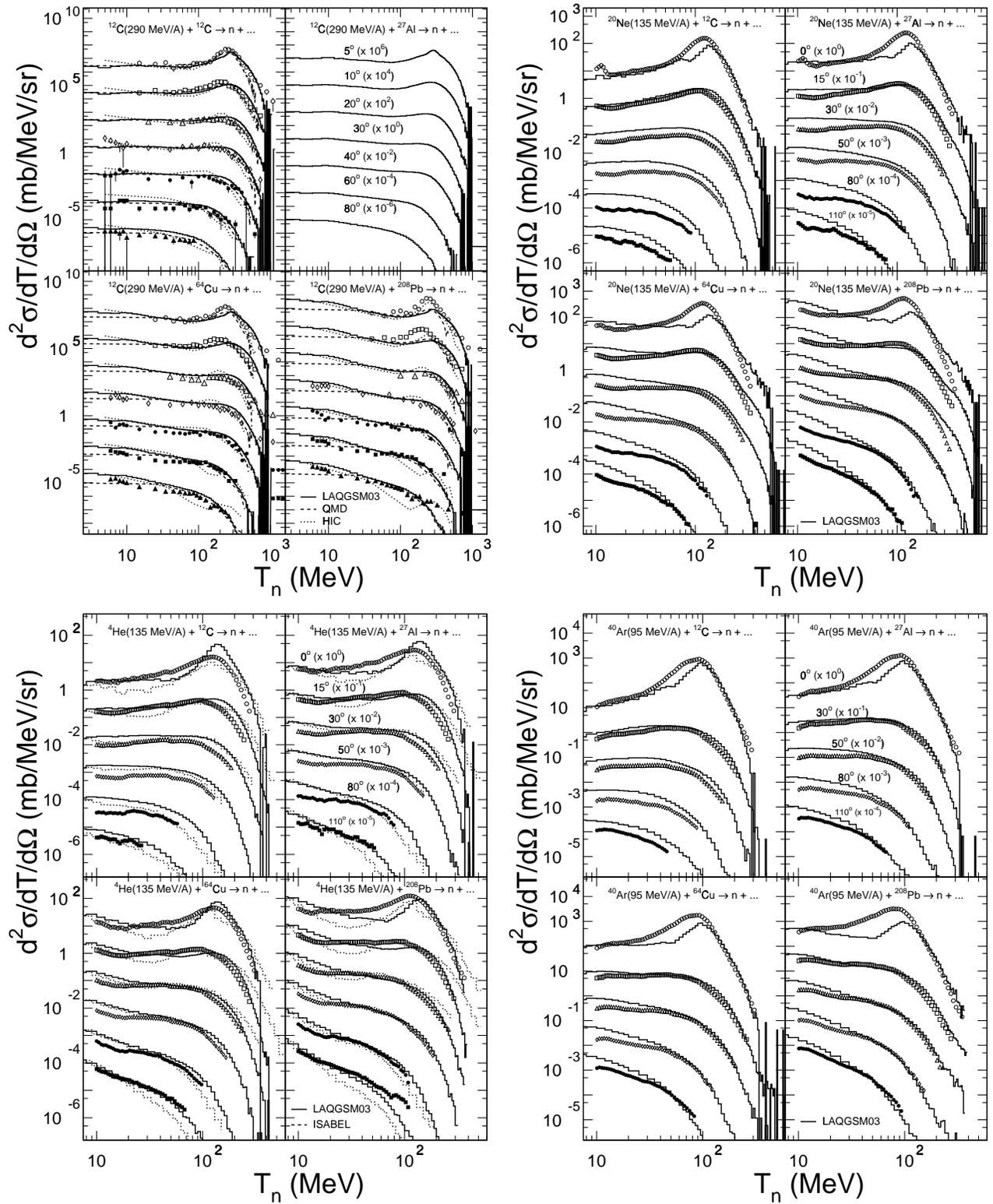

**FIGURE 2.** Comparison of measured double differential cross sections of neutrons from 290 MeV/nucleon $^{12}$C [2], 135 MeV/nucleon $^{20}$Ne [1], 135 MeV/nucleon $^{4}$He [1], and 95 MeV/nucleon $^{40}$Ar [1] on $^{12}$C, $^{27}$Al, $^{64}$Cu, and $^{208}$Pb with our present LAQGSM03 [9] results and calculations by JQMD [3], HIC [4], and ISABEL [6]. Experimental data for 290 MeV/nucleon $^{12}$C on Al are not yet available so we present here only our predictions from LAQGSM03.

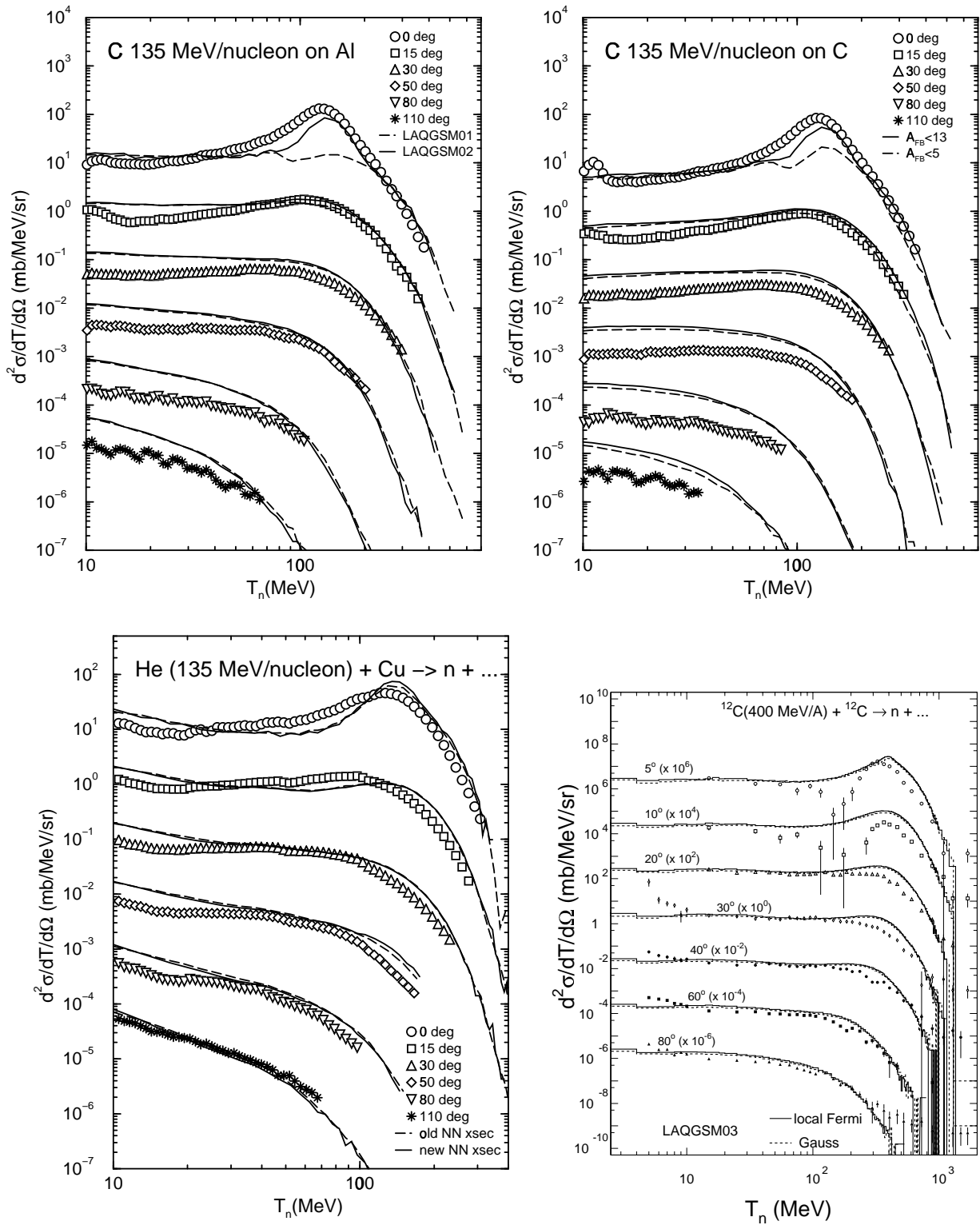

**FIGURE 3.** Analysis of the measured neutron spectra from 135 MeV/nucleon $^{12}$C on $^{27}$Al [1], 135 MeV/nucleon $^{12}$C on $^{12}$C [1], 135 MeV/nucleon $^{4}$He on $^{64}$Cu [1], and 400 MeV/nucleon $^{12}$C on $^{12}$C [2] with different versions of LAQGSM, as described in the text.

We see that on the whole, all four models describe reasonably well most of the measured neutron spectra, although different models agree differently with data from specific reactions and some serious discrepances are observed for some reactions. The biggest disagreement is seen for forward spectra, in the region of the quasielastic peak, and for the low-energy part of the spectra measured at 95 MeV/nucleon; the lower in the energy of the incident beam, the bigger the discrepancy.

We have analyzed with LAQGSM03 possible reasons for some of the observed discrepancies, hoping to outline ways to further improve such models to address these problems. The left upper plot in Fig. 3 shows results from an older version of LAQGSM, marked as LAQGSM01, where the unstable isotopes produced in a reaction do not decay (it was assumed in that version that such decays should be performed by the transport code where LAQGSM01 is used as an event generator). After we decay such unstable isotopes (results marked as LAQGSM02), the agreement of calculations with the data improves considerably. In the right upper plot of Fig. 3, we shows results of LAQGSM02 when we consider Fermi break-up processes instead of preequilibrium+evaporation after the cascade stage of reactions either for nuclei with masses $A < 13$ or for $A < 5$. We see a much better agreement in the former case; therefore we choose the criteria of implementing Fermi break-up for all nuclides with $A < 13$. On the left bottom plot of Fig. 3, we present results of LAQGSM02 using old appproximations for elementary $np$, $pp$, and $nn$ cross sections (marked as "old NN xsec"), and by the newer version of our code, LAQGSM03 [9], that uses new and better approximations for such cross sections. We see that for these particular reactions, the new cross sections provide results very similar to the old ones, although this is not so for all reactions and we get with the new cross sections much better results for certain reactions (see details in [9]). Finally, the right bottom plot of Fig. 3 shows results calculated by LAQGSM03 simulating the nuclear density described by a Fermi distribution, and using a Gaussian parametrization, as is often done in the literature for light nuclei [10]. We see that the results provided by these two nuclear densities are almost the same.

We conclude that using a better Fermi break-up model, better elementary cross sections for $np$, $pp$, and $nn$ interactions in INC simulations, and taking into account the decay of unstable residual nuclei allows us to improve to a certain degree the agreement of calculations with the data, but some serious disagreements still remain for several reactions, especially at low incident energies. The other major discrepancy occurs for 0-degree spectra. More measurements of forward spectra could aid in the improvement of these simulation tools. The discrepancies at low incident energies are probably due to the use of the thin-target assumption in all the models (except for JQMD, where a projectile energy spread due to target thickness was approximately considered in our calculations), while the targets measured in [1, 2] may not be thin enough. We plan to check this hypothesis soon, when all the models used here are available as event generators in transport codes, such as MARS, MCNPX, and PHITS.

This work was partially supported by the US Department of Energy, the Moldovan-US Bilateral Grants Program, CRDF Project MP2-3045, and the NASA ATP01 Grant NRA-01-01-ATP-066.

## REFERENCES

1. Sato, H., et al., *Phys. Rev.*, **C64**, 034607 (2001).
2. Iwata, Y., et al., *Phys. Rev.*, **C64**, 054609 (2001).
3. Niita, K., et al., *Phys. Rev.*, **C52**, 2620–2635 (1995).
4. Bertini, H. W., et. al., ORNL-TM-4134, Oak Ridge (1974).
5. Dresner, L., ORNL-TM-196, Oak Ridge (1962); Guthrie, M. P., ORNL-TM-3119, Oak Ridge (1970).
6. Yariv, Y., and Fraenkel, Z., *Phys. Rev.*, **C24**, 488 (1981).
7. Prael, R. E., and Lichtenstein, H., LANL Report LA-UR-89-3014, Los Alamos (1989).
8. Gudima, K. K., Mashnik, S. G., and Sierk, A. J., LANL Report LA-UR-01-6804, Los Alamos (2001).
9. Mashnik, S. G., et at., Research Note X-5-RN(U)04-08, Los Alamos (2004) and paper #430 in this Proc.
10. Bertulani C. A., and Danielewicz, P., *Introduction to Nuclear Reactions*, IOP Publ. Ltd, Cornwall, UK, 2004.